\documentclass[aps,prb,twocolumn,floatfix]{revtex4-2}

\usepackage{amsmath}

\usepackage{amssymb}
\usepackage{graphicx}
\usepackage{bm}
\usepackage{color}
\usepackage{times}
\begin{document}

\title{Higher-order topology in Fibonacci quasicrystals }

\author{Chaozhi Ouyang$^{1,2}$}
\author{Qinghua He$^{1,2}$}
\author{Dong-Hui Xu$^{3}$}
\email{donghuixu@cqu.edu.cn}
\author{Feng Liu$^{1,2,4}$}
\email{ruserzzz@gmail.com}

\affiliation{$^{1}$Institute of High Pressure Physics, Ningbo
  University, Ningbo, 315-211, China}

\affiliation{$^{2}$School of Physical Science and Technology, Ningbo
  University, Ningbo, 315-211, China}

\affiliation{$^{3}$Deparment of Physcis \& Chongqing Key Laboratory for Strongly Coupled Physics, Chongqing University, Chongqing 400044, China}

\affiliation{$^{4}$Department of Nanotechnology for Sustainable Energy, School of Science and Technology, Kwansei Gakuin University, Gakuen 2-1, Sanda 669-1337, Japan}

\begin{abstract}
In crystalline systems, higher-order topology, characterized by topological states of codimension greater than one, 
typically arises from the mismatch between Wannier centers and atomic sites, leading to filling anomalies.
However, this phenomenon is less understood in aperiodic systems, such as quasicrystals, where Wannier centers are not well defined.
In this study, we examine Fibonacci chains and squares, a quintessential type of quasicrystal, to investigate their higher-order topological properties.
We discover that topological interfacial states, including corner states, can be inherited from their higher-dimensional periodic counterparts, such as the two-dimensional Su-Schrieffer-Heeger model. 
This finding is validated through numerical simulations of both phononic and photonic Fibonacci quasicrystals by the finite element method, revealing the emergence of topological edge and corner states at interfaces between Fibonacci quasicrystals with differing topologies inherited from their parent systems. 
Our results not only provide insight into the higher-order topology of quasicrystals but also open avenues for exploring novel topological phases in aperiodic structures.
\end{abstract}

\maketitle

\section{Introduction}
Over the past two decades, the concept of topology has become a cornerstone in classifying solid-state materials, revealing essential differences even for materials with seemingly identical energy-band structures. 
This paradigm shift, heralded by the topological band theory, has profoundly reshaped our understanding of solid-state physics~\cite{Hasan2010, Qi2011, Bansil2016}.
One of the most striking properties of topological band theory is the bulk-edge correspondence, connecting the bulk topological invariant in crystalline systems to the
emergence of robust interfacial states in finite, topologically distinct samples~\cite{Hatsugai1993, Liang2006, Hwang2019, Bouhon2019, Wang2021}.
In this evolving landscape, the exploration of higher-order topological phases, which deviate from the conventional topological insulators by exhibiting bulk-corner correspondence, has gained prominence in recent years. 
These higher-order topological phases allow for topological states of codimension greater than one, such as corner states, with potential applications in
topological lasers and quantum computing~\cite{Hararieaar2018,Ezawa2018a, Benalcazar2019, Wu2020a, Watanabe2021, Jung2021, Zhang2020c, Zhang2020d, Pahomi2020, Pan2022, Tang2023}.




Significant progress has been made in understanding higher-order topological states in crystalline systems, as evidenced by a wealth of studies. These studies could be exemplified through several models based on filling anomalies induced by the mismatches between Wannier centers and atomic sites~\cite{Liu2017, Benalcazar2017a, Benalcazar2017B, Ezawa2018, Liu2019}.  Protected by various symmetries~\cite{Trifunovic2019, Khalaf2021,Lei2022, Jia2022}, such as point group symmetry~\cite{Song2017, Langbehn2017, Geier2018}, inversion symmetry~\cite{Khalaf2018}, and chiral symmetry~\cite{Okugawa2019,Benalcazar2022}, higher-order topological states are observed not only in solid-state materials~\cite{Schindler2018,Schindler2018a,Yue2019, LiuBing2019, Choi2020, Aggarwal2021, Zhang2019a}, but also in metamaterials, such as photonic~\cite{Xie2018, Xie2019, Chen2019, Mittal2019,Cerjan2020,Wang2021a}, sonic crystals~\cite{Serra-Garcia2018, Xue2019, Xue2019a, Zhang2019, Zhang2019b,Lin2020, Lin2020a}, and electric circuit arrays~\cite{Imhof2018,Liushuo2020,Wangzhu2023}.

A less explored but equally interesting direction is the exploration of these states in systems lacking transnational symmetry, such as quasicrystals. These ideas have recently been explored in two-dimensional (2D) quasicystals~\cite{ChenR20PRL,Varjas19PRL,Hua20PRB,Spurrier20PRR,Huang2022,Wang22PRL}, which requires a fine-tuned mass term to generate a higher-order topology. While these initial studies have shed some light on the interplay between aperiodicity and high-order topology, the fine-tuned mass may hinder the realization of higher-order topology in quasicrystals.
A common approach to probing the topology in quasicrystals involves examining their higher-dimensional parent systems. 
For instance, the topology of a one-dimensional (1D) Fibonacci quasicrystal can be elucidated through its mapping to a 2D periodic ancestor in the Harper model, which can be characterized by the Chern number~\cite{Kraus2012, Jagannathan2021,Panigrahi2022}.

Drawing inspiration from these foundational studies, our research shifts focus to a modified Fibonacci quasicrystal, whose
higher-dimensional ancestor is the 2D Su-Schrieffer-Heeger (SSH) model. 
The 2D SSH model has several distinct topological phases characterized by the vector Zak phase or equivalently the Wannier centers, depending on the ratio between intra-cell and inter-cell hopping. 
In the nontrivial topological phase of the 2D SSH model, filling anomalies give rise to topological corner states in addition to edge states.
By extending the Fibonacci chains to Fibonacci squares, it is possible to explore higher-order topological interfacial states and investigate the preservation of topological properties from high-dimensional parent systems to their lower-dimensional quasicrystal counterparts. 
Such preservation of topological property cross dimensions without translational symmetry can be verified using artificial structures such as the phononic and photonic quasicrystals, which may offer a hint peeping at the topology of various quasicrystals.

In this work, we employ the cut-and-project method to construct modified Fibonacci chains and squares from the 2D SSH model.
We introduce topological interfaces in these structures, derived from parent systems belonging to different topological classes, 
and investigate the emergence of topological edge and corner states.
These states are observed in phononic and photonic quasicrystals through numerical simulations using the finite element method by COMSOL.
Our findings suggest that lower-dimensional systems, even in the absence of translational symmetry, can exhibit topological states influenced by the distinct topological classifications of their higher-dimensional periodic ancestors. This research contributes to a deeper understanding of topology in aperiodic systems and may offer new insights into the design and application of topological materials.

The structure of the remaining parts is organized as follows. In Sec. II, we introduce the modified Fibonacci chains and the corresponding photonic and phononic quasicrystals whose parent system is the 2D SSH model using the cut-and-project method. In Sec. III, we discuss the topological interfacial states formed in photonic and phononic Fibonacci chains. In Sec. IV, we extend the modified Fibonacci chains to Fibonacci squares and explore higher-order topological phases hosting corner states in the phononic and photonic Fibonacci squares. Finally, the discussion and summary are given in Sec. V.  

\section{Construction of modified Fibonacci chains}
The Fibonacci chain, a 1D quasicrystal, is constructed using a recursive relation, $F_n=F_{n-1}+F_{n-2}$.
Starting with $F_1=F_2=1$, $F_{n-1}/F_{n-2}$  asymptotically approaches the golden ratio $\tau=(1+\sqrt{5})/2$ as $n$
becomes large.
In our model, these elements $F_1$ and $F_2$ represent two different types of atomic sites, denoted as letters ``A'' and ``B''. 
These sites differ in terms of spacing distances and resulting hopping amplitudes, similar to the SSH model~\cite{Su1979}.
Applying the Fibonacci sequence, we can construct a simple quasicrystal exemplified by an atomic chain such as "ABBABABBABBA$\cdots$''.

The tight-binding Hamiltonian of the Fibonacci chain can be written as 
\begin{equation}
    \mathcal{H}=\sum_nt_nc^\dagger_{n+1}c_n+\text{H.c.},    
\end{equation}
where $t_n=(t+\lambda V_n)$ denotes the hopping amplitude between site $n$ and $n+1$, 
$V_n=2[\lfloor (n+2)/\tau \rfloor)-\lfloor (n+1)/\tau \rfloor)]-1$ is the characteristic function with $\lfloor x\rfloor$ the floor function, and $\lambda$ is the hopping difference between A and B sites. 
It is noted that the characteristic function can be equivalently obtained by the cut-and-project method, where the higher-dimensional parent system is a simple 2D square lattice for the conventional Fibonacci chain~\cite{Jagannathan2021}.

In the following, we extend the conventional Fibonacci chain to a modified one through the cut-and-project method, whose higher-dimensional parent system is the 2D SSH model, which has several distinct topological phases identified by the Wannier centers. 
We expect these modified Fibonacci chains to inherit topological properties including higher-order ones from their parent systems, and we verify this observation through numerical simulations of phononic and photonic quasicrystals.

\subsection{cut-and-project method}

\begin{figure}[t]
\leavevmode
\begin{center}
\leavevmode
\includegraphics[clip=true,width=0.99\columnwidth]{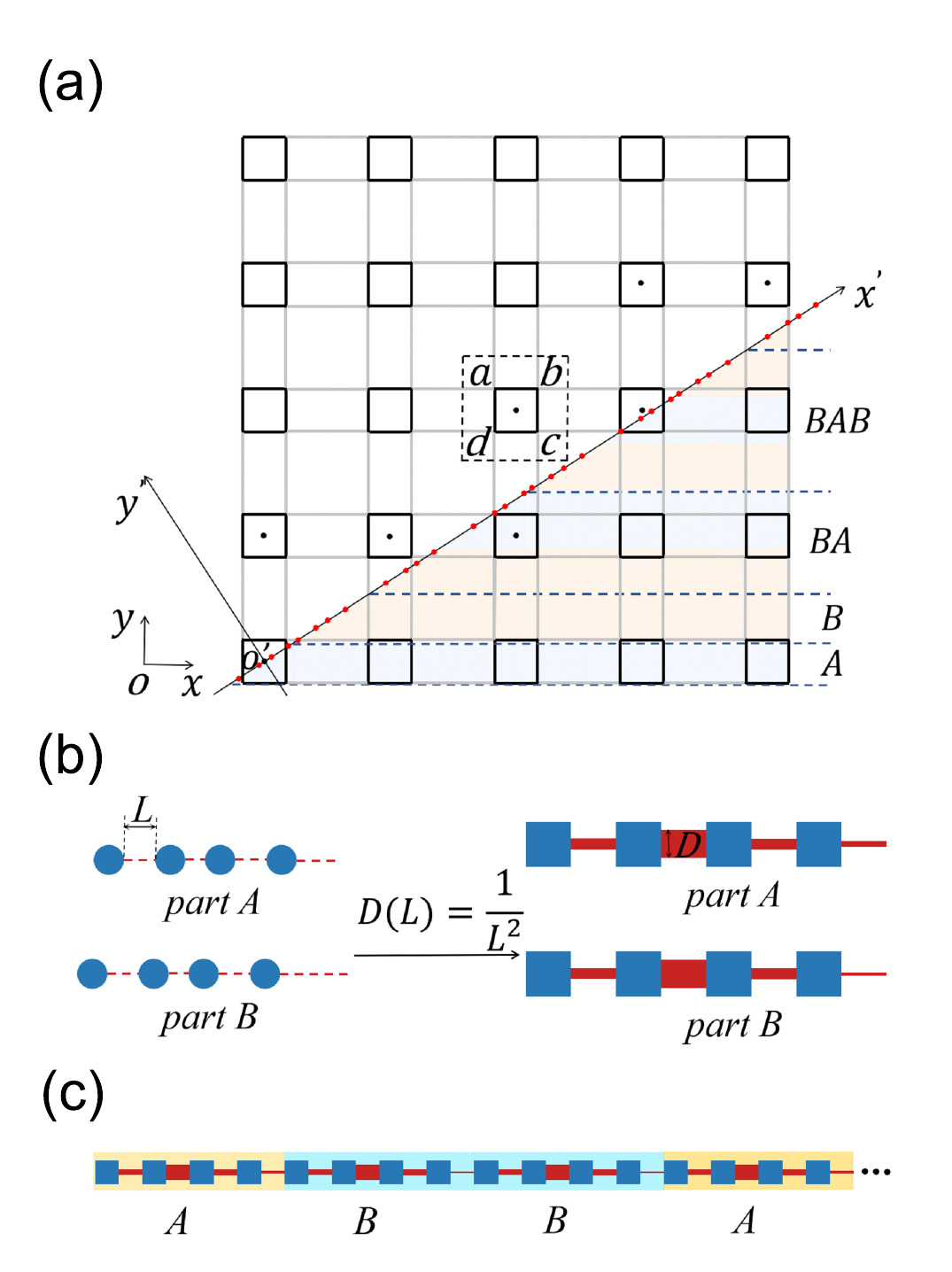}
\end{center}
\caption{(a) Schematic of the cut-and-project for the modified Fibonacci chain. The higher-dimensional parent system is the 2D SSH model, which has four sites in one unit cell. When the projection angle is $\tan\theta=1/\tau$, Fibonacci sequence forms with parts A and B differed by the distances, i.e., $(l+d)\cos\theta$ and $(l+d)\sin\theta$. 
(b) Parts A and B of the modified Fibonacci chain and the corresponding phononic quasicrystal structure. The phononic modified Fibonacci chain is constructed by mapping the distance between sites to the width of coupling waveguide as $D(L)=1/L^2$.
(c) Illustration of the modified phononic Fibonacci chain. 
}
\end{figure}

Construction of the Fibonacci chain using the cut-and-project method provides an intuitive approach to understanding its structure.
As illustrated in Fig.~1(a), this method starts with a parent 2D square lattice. 
The Fibonacci quasicrystal is then generated by projecting these lattice sites onto an axis labeled $x^\prime$.
The orientation of this axis is critical, defined by an angle $\theta$ such that $\tan\theta=\omega$, where $\omega=1/\tau$.
The projection strip used for this operation has a width equal to one unit cell of the lattice.
As a result, the spacing between the projected centers of the unit cells can take on two distinct values:
$\cos\theta=1/\sqrt{1+\omega^2}$ and $\sin\theta=\omega/\sqrt{1+\omega^2}$.
These spacing values correspond to the A and B sites of the Fibonacci chain, as previously mentioned.

Incorporating the 2D SSH model into the cut-and-project method, we obtain a modified Fibonacci chain.
The topological phases of the 2D SSH model are distinguished by two tpes of hopping: 
inter-cell hopping $\gamma_\text{inter}$ and intra-cell hopping $\gamma_\text{intra}$.
As illustrated in Fig.~1(a), these hoppings are represented with varying distances and line thicknesses.
Crucially, the 2D SSH model exhibits four distinct topological phases, which can be characterized by the vector Zak phase, or equivalently, by the Wannier center coordinates $(\nu_x,\nu_y)$, where $\nu_i$ can be either 0 or 1/2 in terms of the lattice constant along the $i$-direction.
For the symmetric case under the point group $C_4$ that we focus on, $\nu_x=\nu_y=0$ for $\gamma_\text{intra}>\gamma_\text{inter}$, and $\nu_x=\nu_y=1/2$ for $\gamma_\text{intra}<\gamma_\text{inter}$. 
The interplay between these hopping parameters in the 2D SSH model directly influences the topological characteristics of the resulting modified Fibonacci chain, providing a rich framework for exploring topological phenomena in quasicrystalline structures.

In the 2D SSH model, each unit cell contains four sublattices. 
By projecting these sublattices onto the $x^\prime$ axis, as shown by the red dots in Fig.~1(a), 
we derive the modified Fibonacci chain. 
This projection results in two distinct types of tiling within the chain, labeled as parts A and B, similar to the conventional Fibonacci chain.
However, a key difference in the modified version is that each of these parts encompasses four sites.  
The positions of these four sites along the  $x^\prime$ axis are determined by the following euqations: 
\begin{equation}
\begin{split}
x^\prime_a=(nl+\frac{l}{2})\sin\theta+(ml-\frac{l}{2})\cos\theta+md\cos\theta+nd\sin\theta\\
x^\prime_b=(nl+\frac{l}{2})\sin\theta+(ml+\frac{l}{2})\cos\theta+md\cos\theta+nd\sin\theta\\
x^\prime_c=(nl-\frac{l}{2})\sin\theta+(ml+\frac{l}{2})\cos\theta+md\cos\theta+nd\sin\theta\\
x^\prime_d=(nl-\frac{l}{2})\sin\theta+(ml-\frac{l}{2})\cos\theta+md\cos\theta+nd\sin\theta,   
\end{split}
\end{equation}
where $(m,n)$ represents the index of the unit cell in the 2D SSH model, 
$a,b,c,d$ are the labels of the sublattices as indicated in Fig.~1(a),
$l$ is the distance between sites within a unit cell,
and $d$ is the distance between adjacent unit cells.
The values of $(m,n)$ are constrained by the relation:
\begin{equation}
     \frac{n-1}{m+1} < \tan\theta \leq \frac{n}{m}.
\end{equation}
It is important to note that for $d>l$, there is a possibility for the sites of neighboring parts A and B to overlap,
a factor that must be considered in analyzing the structure of the modified Fibonacci chain.

\subsection{modified phononic and photonic Fibonacci quasicrystals}
We translate the theoretical framework of the cut-and-project method from the parent 2D SSH model into tangible constructs of phononic and photonic Fibonacci chains. 
Starting with the photonic quasicrystal, we follow a straightforward method:
artificial atoms, such as dielectric rods, are positioned in accordance with the projected sites on the $x^\prime$ from the 2D SSH model as indicated by Eq.~(2).
This arrangement is depicted in the left panel of Fig.~1(b).
In our numerical simulations, we set the radius of the dielectric rod at $0.2~\text{[mm]}$ and the dielectric constant at 100 on an air background. 
We employ a perfect electric conductor as the boundary condition in the simulation of photonic quasicrystals. Notably, because of the absence of a characteristic length in Maxwell's equations, the size of the photonic quasicrystal can be adjusted without affecting its fundamental properties. The parameters selected here serve illustrative purposes only. This approach effectively replicates the modified Fibonacci chain, encapsulating its long-range interactions, and allowing us to explore its topological properties in photonic analogues.

For the phononic quasicrystal, we use square resonators connected by waveguides. 
The coupling between each resonator is proportional to the width of the waveguides, and we map the distance between the sites to the width of the waveguide as $D=1/L^n$ with $n=2$. It should be noted that for other mapping functions, the simulation results remain qualitatively similar. For our simulations, we set the dimensions of each square resonator to 30~[mm] in length, with the waveguides also at a length of 30~[mm]. The uniformity of waveguide length ensures consistent propagation characteristic throughout the whole phononic quasicrsytals. 
The boundary condition applied in the simulation of phononic quasicrystals is the hard boundary condition, which ensures that the sound pressure is zero at the boundaries, mimicking an isolated system. Figure 1(c) illustrates an example of the modified phononic Fibonacci quasicrystal, where the various shades are used to denote parts A and B.

\section{topological interfacial states}
In this section, we delve into the phenomenon of topological interfacial states, which arise at the junction of two topological distinct systems.
Our focus is on the demonstration of the emergence of these topological interfacial states in modified Fibonacci chains derived from phononic and photonic quasicrystals, each with distinct topological characteristics inherited from parent systems.
This demonstration is intriguing because, in quasicrystals, the concept of a Wannier center is not well defined because of the absence of translational symmetry, unlike the case of the SSH model. 

In addition to demonstration of the emergence of topological interfacial states, symmetry protection of topological states in quasicrystals is also important. 
Like in the SSH model, chiral symmetry pins the energy of edge states at zero.
The role of chiral symmetry in these quasicrystalline systems presents a rich area for exploration, potentially revealing new insights into symmetry protection in topological states. An in-depth investigation of chiral-symmetry protection in the topological interfacial states of quasicrystals is reserved for future work.


\begin{figure}[t]
\leavevmode
\begin{center}
\leavevmode
\includegraphics[clip=true,width=0.99\columnwidth]{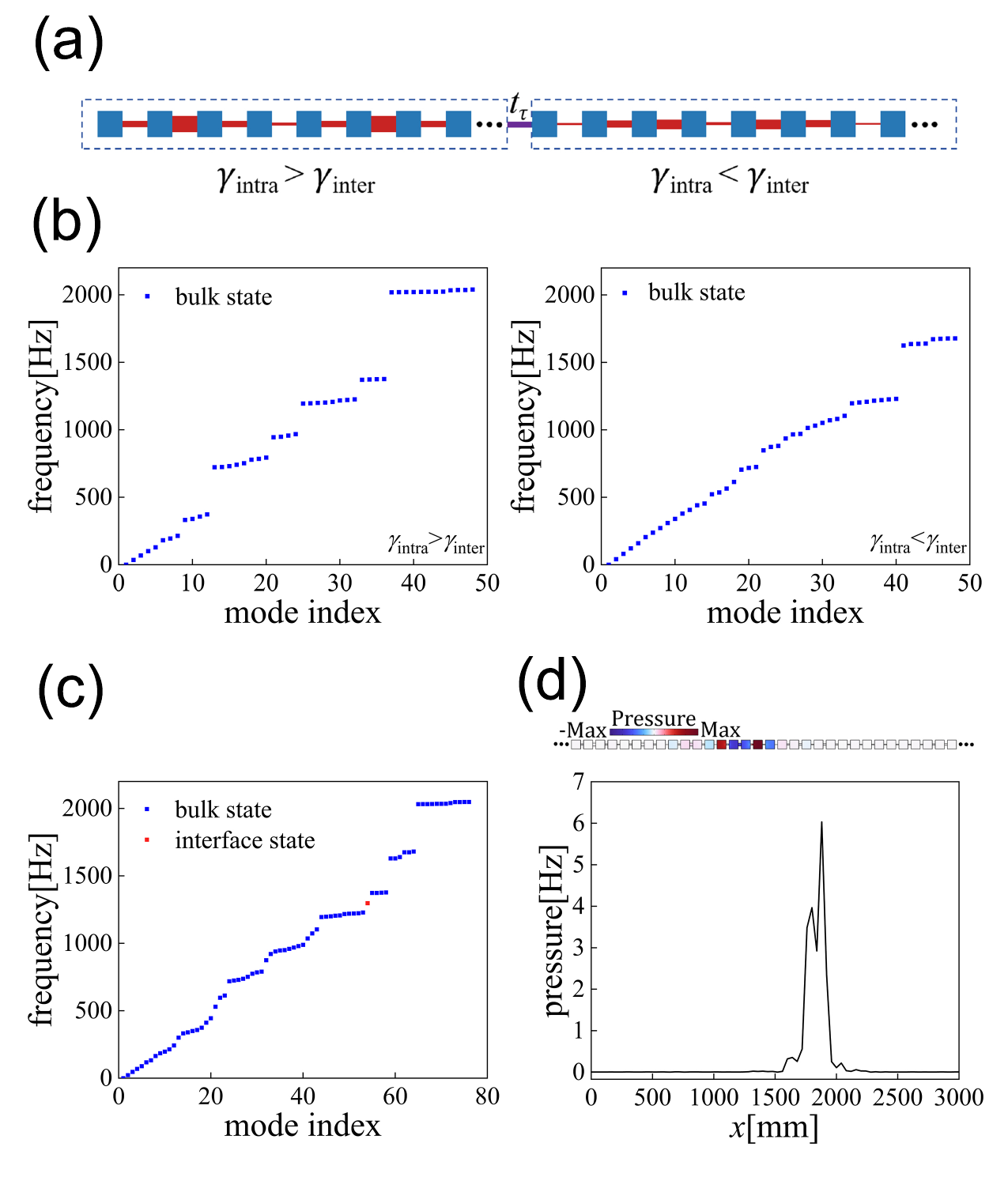}
\end{center}
\caption{(a) Topological interface formed by the two types of modified phononic Fibonacci chains, whose higher-dimensional parent systems belonging two topological classes of the 2D SSH model. The coupling between two chains is denoted by $t_\tau$ and controlled by the width of coupling waveguide.
(b) Eigenspectra of the two types of modified phononic Fibonacci chains in (a), as displayed by left and right panels, respectively.
(c) Eigenspectra of the combined structure of the two types of modified phononic Fibonacci chains.
(d) Wave profile of the topological interfacial state as sound pressure in the setup of (a). 
}
\end{figure}

\subsection{phononic topological interfacial states}
We construct a topological interface by joining two phononic Fibonacci chains, each originating from the 2D SSH model but with distinct topological characteristics.
This interface, as illustrated in Fig.~2(a), servers as a physical realization of the boundary between different topological phases of the quasicrystals.
The left phononic chain is adopted from the 2D SSH model with $l=1\text{[mm]}$ and $d=3~\text{[mm]}$, 
effectively resulting in a scenario of $\gamma_\text{intra}>\gamma_\text{inter}$.
On the contrary, the right phononic chain adopts the parameters $l=3~\text{[mm]}$ and $d=~1\text{[mm]}$, which correspond to $\gamma_\text{intra}<\gamma_\text{inter}$.

In Fig.~2(b), we display the eigenspectra of the two topologically distinct types of phononic Fibonacci chains derived from the 2D SSH model.
There appear multiple gaps in the eigenspectra similar to the conventional Fibonacci chain defined in Eq.~(1). 
The contrast in the eigenspectra of the modified Fibonacci chain is the four subbands in the middle, which reflect the underlying structure of the four sublattices in the 2D SSH model, as displayed by the left panel of Fig.~2(b).
In the 2D SSH model, it is observed that the two different topological phases can be exchanged by redefining the unit cell following a half-period shift.
This shift results in a pattern mapping between the two topologically distinct Fibonacci chains, where part A is transformed into part B and part B becomes BA.
However, despite this transformation, parts A and B in the two chains remain nonequivalent due to the different values of $l$ and $d$, which do not result in the closure of the band gap in their parent systems.

As displayed in Fig.~2(c), we observe the emergence of a topological interfacial state within the shared gap of the two phononic Fibonacci chains.
Up to the frequency that we solve, there is only one topological interfacial state.
By adjusting the coupling strength, denoted as $t_\tau$, between the two chains, we can effectively modulate the frequency of this state within a certain range. This tunability offers a versatile tool for manipulating the properties of the interfacial state for potential applications.

The wave profile of the topological interfacial state, as shown in Fig. 2(d), reveals another critical feature: the state exhibits expected exponential decay along both directions from the interface. 
This decay is a hallmark of topological interfacial states, indicating the localized nature of these states at the interface of the two distinct topological phases.

The total number of sites in parts A and B of the simulation is 48, which corresponds to the Fibonacci sequence ``ABBABABBABBA''. 
For a larger number of sites, the eigenspectra remain qualitatively the same, and the band gaps do not close. 

\begin{figure}[t]
\leavevmode
\begin{center}
\leavevmode
\includegraphics[clip=true,width=0.99\columnwidth]{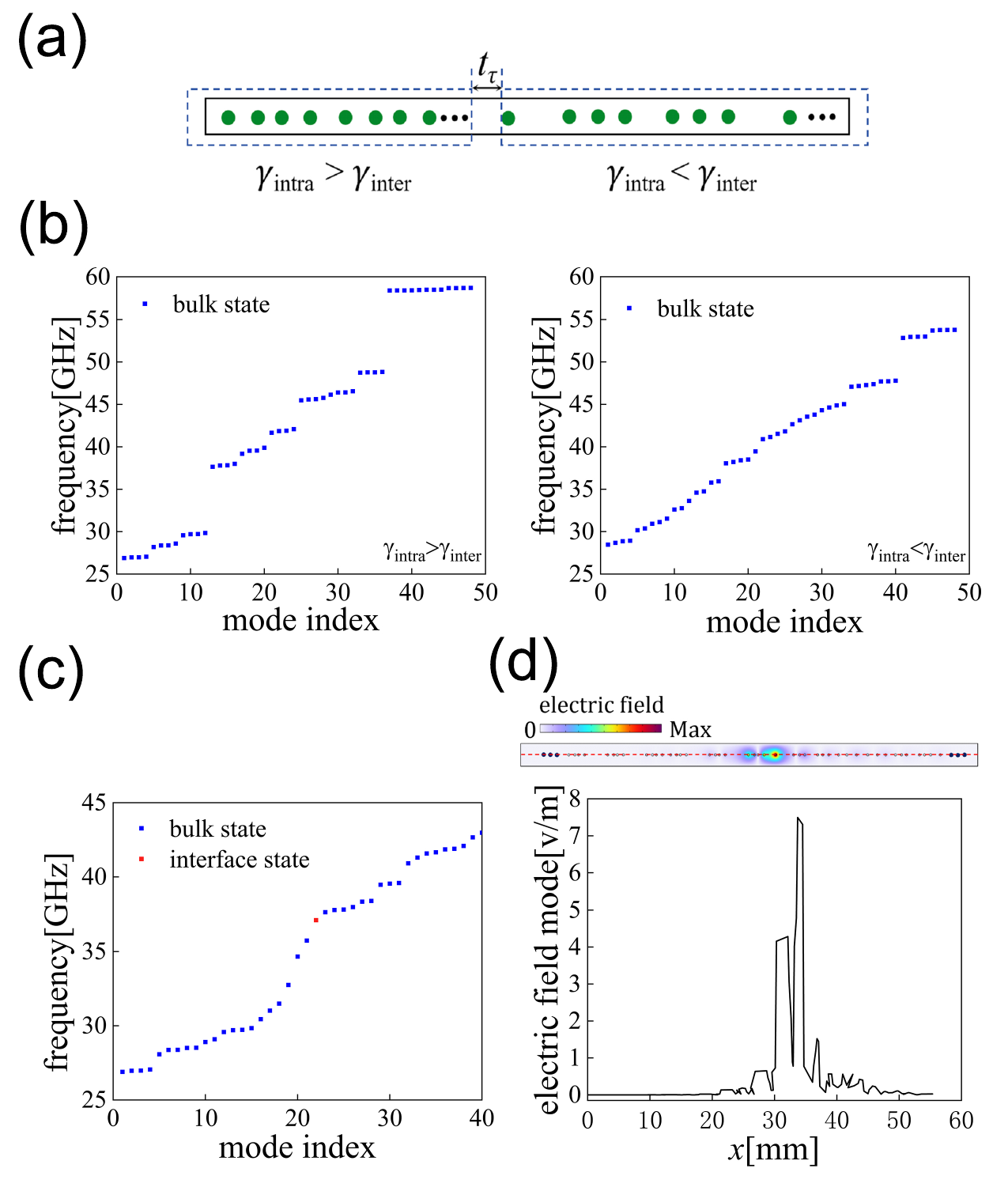}
\end{center}
\caption{ (a) Topological interface formed by the two types of modified photonic Fibonacci chains, whose higher-dimensional parent systems belonging two topological classes of the 2D SSH model. The coupling between two chains is denoted by $t_\tau$ and controlled by the distance between two chains.
(b) Eigenspectra of the two types of modified photonic Fibonacci chains in (a), as displayed by left and right panels, respectively.
(c) Eigenspectra of the combined structure of the two types of modified photonic Fibonacci chains.
(d) Wave profile of the topological interfacial state as electric field in the setup of (a).  
}
\end{figure}

\subsection{photonic topological interfacial states}

In addition to the phononic Fibonacci chains, it is equally compelling to investigate the topological interfacial states in photonic quasicrystals.
One notable aspect of the photonic case is the non-negligible impact of long-range interactions.

Figure 3(a) depicts the topological interface between the two distinct types of photonic Fibonacci chains, 
represented by green dots indicating the placement of dielectric rods.
Upon examination of the eigenspectra of these photonic chains, as shown in Fig.~3(b), we find a similarity to the phononic case, suggesting similar topological behaviors.
The eigenspectra of these two types of photonic Fibonacci chains are displayed in Fig.~3(b), which looks similar to the phononic case. 
The eigenspectra of the combined structure of these photonic Fibonacci chains, as displayed in Fig. 3(c), reveal a topological interfacial state, marked by a red dot.
Moreover, the wave profile of this topological interfacial state, represented as an electric field in photonic Fibonacci chains, is displayed in Fig. 3(d).

\begin{figure}[t]
\leavevmode
\begin{center}
\includegraphics[clip=true,width=0.99\columnwidth]{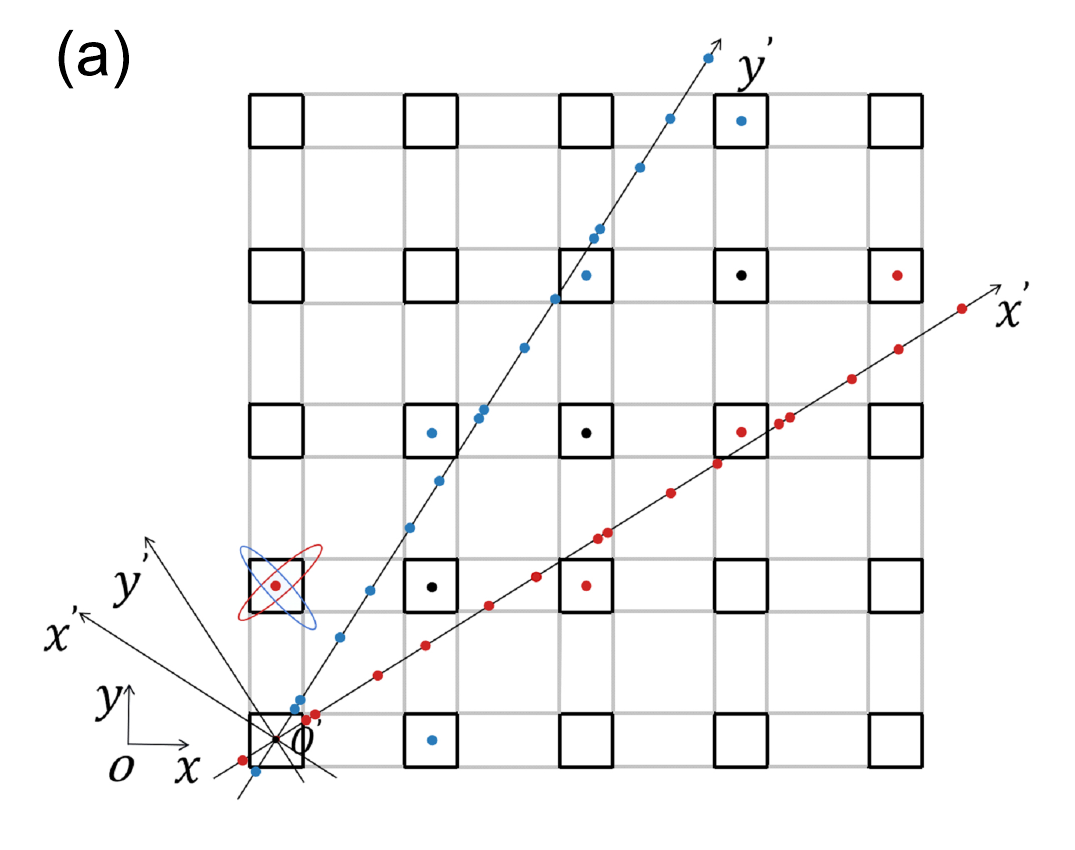}
\end{center}
\caption{Schematic of the extended cut-and-project method for the modified Fibonacci square. There are two projecting axes $x^\prime$ and $y^\prime$, whose projection angles are complementary angles. The diagonal sites of the 2D SSH model are projected to the $x^\prime$ axis, and the off-diagonal sites are projected to the $y^\prime$ axis.   
}
\end{figure}

\begin{figure}[t]
\leavevmode
\begin{center}
\includegraphics[clip=true,width=0.94\columnwidth]{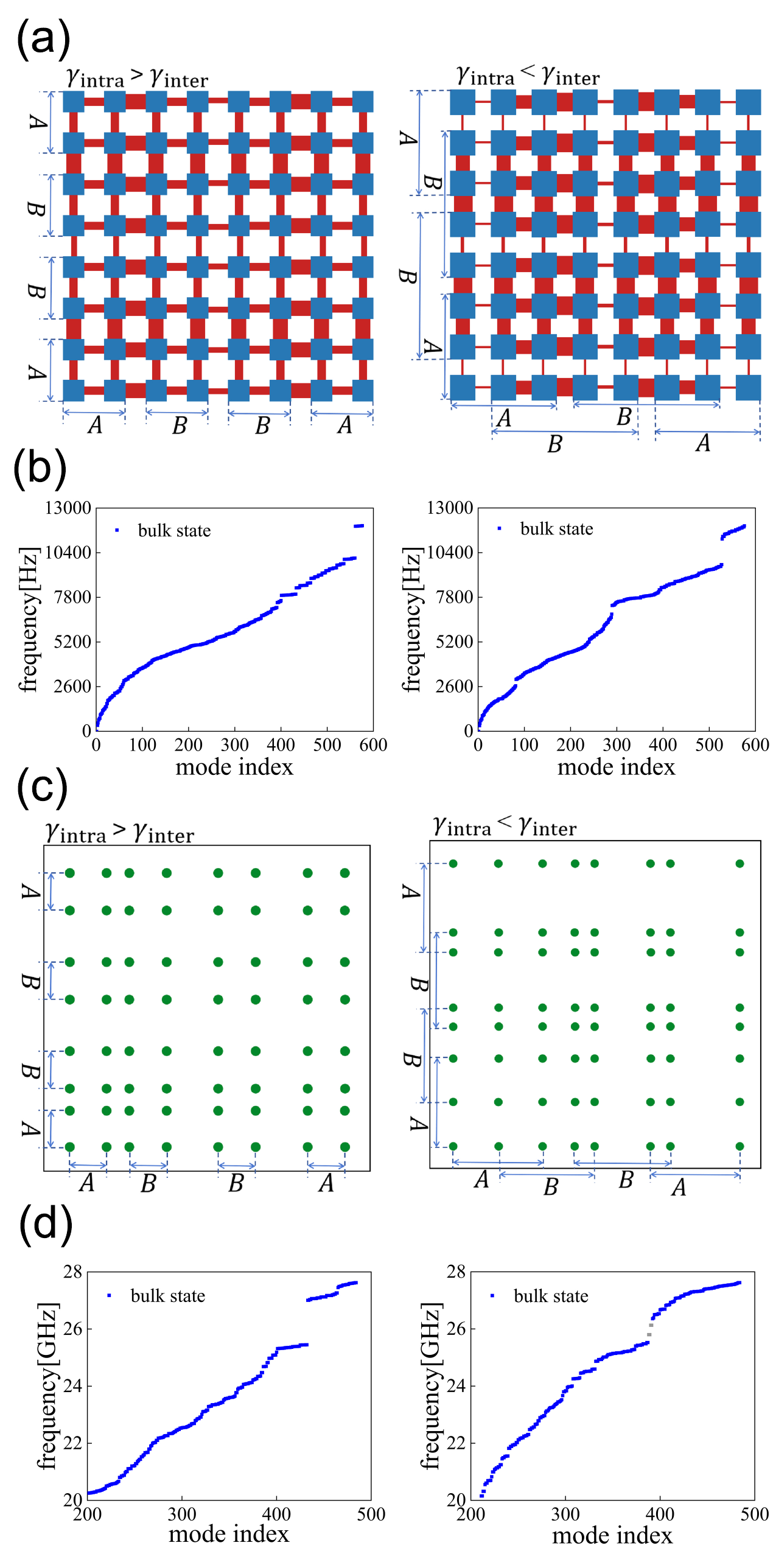}
\end{center}
\caption{(a) Schematics of the two types of modified phononic Fibonacci squares, which belongs two topologically distinct classes of the 2D SSH model, constructed by the extended cut-and-project method. 
(b) Eigenspectra of the two types of modified phononic Fibonacci squares. 
Left panel is for the case of $\gamma_\text{intra}>\gamma_\text{inter}$, and right panel is for  $\gamma_\text{intra}<\gamma_\text{inter}$.
(c) Similar to (a) for the modified photonic Fibonacci squares.
(d) Similar to (b) for the modified photonic Fibonacci squares.}
\end{figure}

\begin{figure}[t]
\leavevmode
\begin{center}
\includegraphics[clip=true,width=0.95\columnwidth]{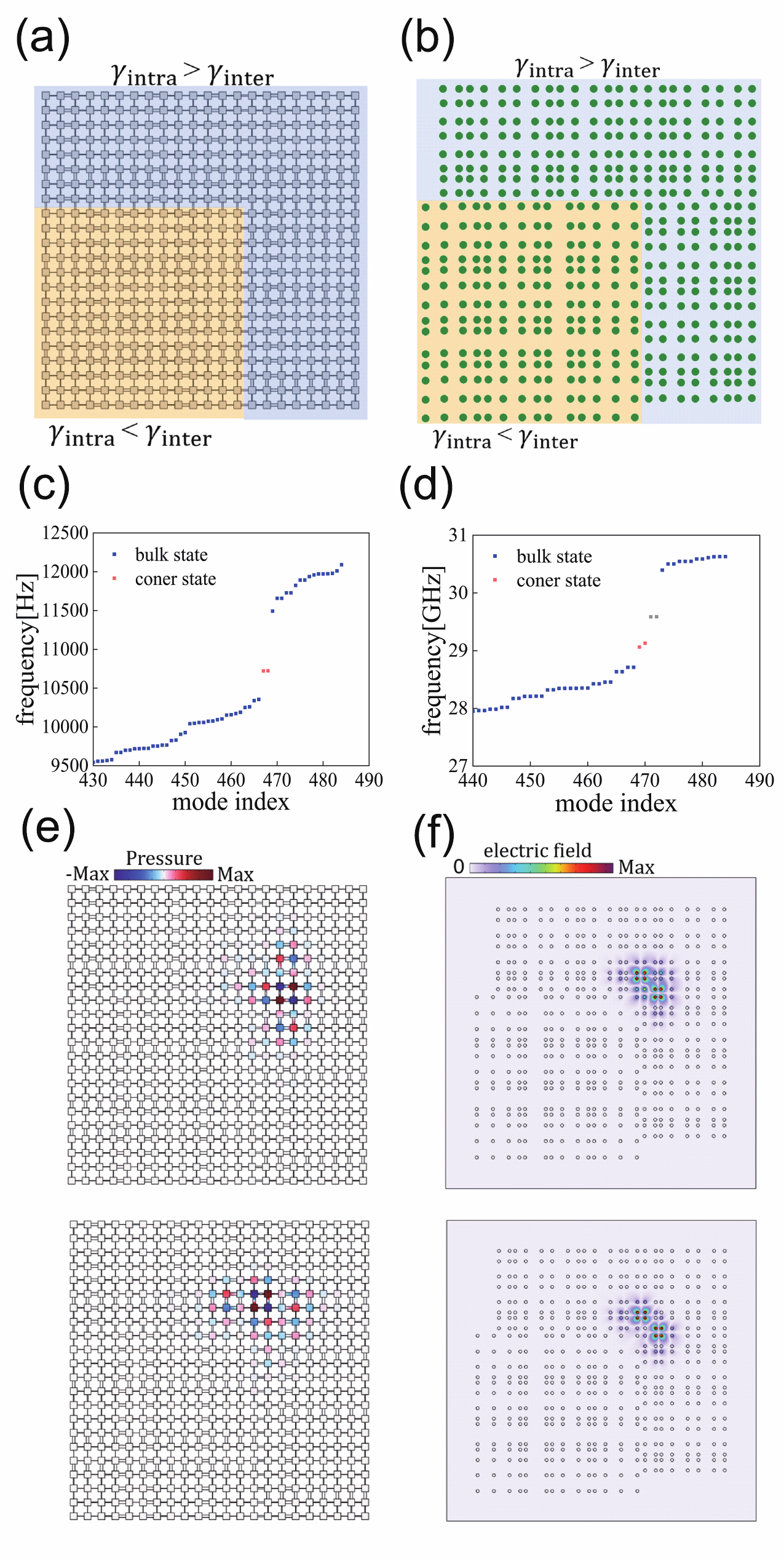}
\end{center}
\caption{(a) Topological interface with a corner structure formed by two types of modified phononic Fibonacci squares.
(b) Topological interface with a corner structure formed by two types of modified photonic Fibonacci squares.
(c) Eigenspectra for the combined structure in (a).
(d) Eigenspectra for the combined structure in (b).
(e) Wave profiles as sound pressure of the topological corner states in (c).
(f) Wave profiles as electric field of the topological corner states in (d).
}
\end{figure}

\section{higher-order topological interfacial states}
In Sec. II(A), we introduce the cut-and-project method for constructing modified Fibonacci chains from the 2D SSH model.
Building upon this method, we extend its application from constructing Fibonacci chains to Fibonacci squares by delving into the study of higher-order topological interfacial states in both phononic and photonic quasicrystals.

\subsection{Construction of modified Fibonacci squares}

In our quest to construct the Fibonacci square, the traditional chain product method proves to be somewhat cumbersome, resulting in a complex structure with 16 sites in each of parts A and B.
To streamline this process, we have adapted the cut-and-project method to project sites on both the $x^\prime$ and $y^\prime$ axes.
This adaptation involves projecting diagonal sites of the 2D SSH model, such as $a$ and $c$ on the $x^\prime$ axis, while aligning the off-diagonal sites, namely $b$ and $d$ on the $y^\prime$ axis.
The schematic of the extended cut-and-project method for modified Fibonacci squares is depicted in Fig.~4, where there are two projection axes, namely $x^\prime$ and $y^\prime$. 
A key aspect of this method is the relationship between the projection angles that satisfies
$\theta_{x^\prime}+\theta_{y^\prime}=\pi/2$.
This angular arrangement is critical, as it imparts symmetry of the $C_4$ point group to the resulting modified Fibonacci square, which simplifies the construction of corresponding phononic and photonic Fibonacci squares. 

\subsection{phononic and photonic modified Fibonacci squares}

Upon determining the projected positions of the sites on the $x^\prime$ and $y^\prime$ axes, we construct the corresponding modified phononic and photonic Fibonacci squares. 
Figure 5(a) displays the structures for the two different topological phases in the phononic case.
It is seen that in modified phononic Fibonacci squares, Fibonacci sequences appear along the $x-$ and $y-$ directions.
In the nontrivial phase that $\gamma_\text{intra}<\gamma_\text{inter}$, 
we notice a mix of sites for parts A and B similar to the modified Fibonacci chain.
The length of the squares and the coupling waveguide is $5~\text{[mm]}$. 
The mapping function between the distance and the width of the coupling waveguide is $D=1/L^2$.

Figure 5(b) displays the eigenspectra for the modified phononic Fibonacci squares.
There is a common band gap for the two topologically distinct modified Fibonacci squares around $10^4~\text{[Hz]}$, allowing us to explore the higher-order topological states. 

Turning to the photonic counterpart, Fig. 5(c) illustrates the modified photonic Fibonacci squares, with the left panel representing the topologically trivial case and the right panel depicting the nontrivial case.
The radius of the dielectric robs is $0.2~\text{[mm]}$, and the dielectric constant is 100.
The remaining part is air.
The eigenspectra of the modified photonic Fibonacci squares is displayed in Fig.~5(d). Similarly to the phononic case, we observe a common band gap for both types of squares. It is important to note that in the nontrivial photonic Fibonacci squares, boundary effects lead to the emergence of localized states within the band gap.

\subsection{corner states in modified Fibonacci squares}
We demonstrate that topological corner states emerge when two distinct types of modified Fibonacci squares are combined. 
The setup of combined modified phononic and photonic Fibonacci squares are displayed in Figs.~6(a) and 6(b), respectively.
The different shades indicate their topological classifications. 
In each setup, a corner structure is constructed in the combined phononic and photonic Fibonacci squares.

The eigenspectra of these combined structures for both phononic and photonic cases are displayed in Figs. 5(c) and~5(d). 
A notable observation is the appearance of corner states within the common band gaps of these configurations. 
Interestingly, these corner states exhibit double degeneracy, distinguishing them from the topological interfacial states observed in the modified Fibonacci chain. 
This characteristic is further elucidated in the wave profiles of the corner states for both the phononic and photonic cases, shown in Figs. 5(e) and 5(f). 
Here, we observe that, unlike the corner states typically seen in the 2D SSH model, which are localized at the corner sites, the corner states in our modified Fibonacci square model manifest around the corner sites rather than directly at them. 
This unique localization pattern of the corner states in our model provides new insights into the behavior of topological states in quasicrystalline systems and contributes to our understanding of their topological characteristics.

\section{Summary}

In summary, we have proposed a modified Fibonacci quasicrystal whose higher-dimensional parent system is the two-dimensional Su-Schrieffer-Heeger model.
By constructing the corresponding phononic and photonic quasicrsytals of the modified Fibonacci chains and squares, we observe that topological interfacial states
including higher-order one such as corner states emerge in the finite-element simulations.
Our results suggest that, in addition to the Chern-insulator states, higher-order topological states induced by filling anomalies can also be inherited from the parent systems in quasicrystals.
In particular, we observe the double degeneracy of the corner states.
The discussed phononic and photonic quasicrytals of modified Fibonacci quasicrystals offer a platform for further study of the higher-order topological properties in quasicrystals.


 \section*{Acknowledgments}
This work is supported by the NSFC (under No. 12074205, No. 12074108, and No. 12347101), NSFZP Grant No. LQ21A040004, the Natural Science Foundation of Chongqing (Grant No. CSTB2022NSCQ-MSX0568) and the Fundamental Research Funds for the Central Universities (Grant No. 2023CDJXY-048).


\bibliography{references}

\end{document}